\documentclass[10pt,twocolumn]{IEEEtran}
\usepackage{amsmath,amssymb,mathrsfs}
\usepackage{color,array,times}
\usepackage{epsfig,epstopdf,textcomp}
\usepackage{epsfig, subfigure, color}
\usepackage{float, multirow}
\usepackage{amsmath}
\usepackage{wrapfig}
\usepackage{graphicx}
\usepackage{epsfig, subfigure, color}
\usepackage{float, multirow}
\usepackage{epstopdf}
\DeclareGraphicsExtensions{.eps}
\usepackage{amssymb}
\usepackage{algorithm}
\usepackage[noend]{algpseudocode}
\usepackage{tikz}
\usetikzlibrary{matrix}
\usepackage{cancel}
\usepackage{amsmath}
\begin{document}

\title{Study of Design of Rate-Compatible Polar Codes Based on Non-Uniform Channel Polarization}

\author{Robert M. Oliveira and Rodrigo C. de Lamare
\thanks{Robert M. Oliveira and Rodrigo C. de Lamare - Centre for Telecommunications
Studies (CETUC), Pontifical Catholic University of Rio de Janeiro
(PUC-Rio), Rio de Janeiro-RJ, Brasil, E-mails: rbtmota@gmail.com e
delamare@cetuc.puc-rio.br. Part of this work has been reported in a
conference paper presented at ISWCS 2019.}}

\maketitle



\begin{abstract}
We propose a novel scheme for rate-compatible arbitrary-length polar code construction for the additive white Gaussian noise (AWGN) channel. The proposed scheme is based on the concept of non-uniform channel polarization. The original polar codes can only be designed with code lengths that are powers of two. Puncturing, shortening and extension are three strategies to obtain arbitrary code lengths and code rates for polar codes. There are other ways to design codes with arbitrary length but which have encoding and decoding with higher complexity such as multi-kernel, concatenated codes and specific constructions for Belief propagation (BP) or Successive Cancellation (SC) decoding. In general, the quality of the projected bit channels by these arbitrary-length techniques differs from that of the original bit channels, which can greatly affect the performance of the constructed polar codes. The proposed Non-Uniform Polarization based on  Gaussian Approximation (NUPGA) is an efficient construction technique for rate-compatible arbitrary-length polar codes, which chooses the best channels (i.e., selects the positions of the information bits) by re-polarization of the codeword with desired length. A generalization of the Gaussian Approximation is devised for both polarization and re-polarization processes. We also present shortening and extension techniques for design polar codes. Simulations verify the effectiveness of the proposed NUPGA designs against existing rate-compatible techniques.

\end{abstract}


\section{Introdution}
Polar codes, introduced by Arikan \cite{Arikan}, were proved to achieve the symmetric capacity of binary input symmetric discrete memoryless channels (B-DMCs) under a Successive Cancellation (SC) decoder as the codelength goes to infinity. The block error rate converges to zero as the code length $N$ goes to infinity. Polar codes are based on the phenomenon of channel polarization. The channel polarization theorem states that, as the codelength $N$ goes to infinity,  the polarized bit-channels become either a noiseless channels or pure noise channels. The information bits are transmitted over the noiseless bit-channels and the pure noise bit-channels are set to zero (frozen bits). The construction of a polar code involves the identification of channel reliability values associated to each bit to be encoded. This identification can be performed for a code length and a  specific signal-to-noise ratio. Therefore, polar code construction refers to selecting an appropriate frozen bit position pattern and several polar code construction algorithms have been proposed. Polar codes were recently selected by the 3GPP group as the channel codes for the upcoming 5th generation mobile communication standard (5G) uplink/downlink control channel \cite{5G}.

In the construction of polar codes, we recall that the code length $N$ of standard polar codes is limited to powers of two, i.e., $N = 2^n$, however, the code length flexibility is required for practical applications. There are several construction techniques applied in the standard polar code model proposed by Arikan \cite{Arikan} considering the SC decoder. Among the most well-known polar codes construction techniques are the  Bhattacharyya-based approach, proposed by Arikan \cite{Arikan}, the Density Evolution (DE) schemes of Mori \cite{Mori1}\cite{Mori2} and Tal \cite{Tal1}, the Gaussian Approximation (GA) technique of \cite {Trifonov} and the Polarization Weight (PW) algorithm \cite{RZhang}. The Bhattacharyya parameter-based approach was proposed along with Monte-Carlo (MC) simulations to estimate bit channel reliabilities \cite{Arikan}. The MC method needs to estimate the average error probability of each bit-channel via simulations. Therefore, encoding and SC decoding are required in each simulation. In the DE method, we need to compute the probability density function (PDF) of the log-likelihood-ratio (LLR) of each channel first and then choose the channels that are most likely to be correct. Since DE includes function convolutions, its precision is limited by the complexity. Similar to the MC method, the DE method has high computational cost. Extending the ideas of \cite{Mori1} and \cite{Mori2}, Tal \cite{Tal1} devised two approximation methods by which one can get upper and lower bounds on the error probability of each bit-channel using efficient schemes of degrading and upgrading quantization. A bit-channel reliability estimation method for  AWGN channels based on DEGA has been proposed in \cite{Trifonov}, giving accurate results with a significant reduction in complexity if compared with DE. The PW algorithm \cite{RZhang} is a recent attempt to study the partial order (PO) \cite{Schurch} for designing universal polar codes, which performs as well as GA but with much lower complexity. An important  characteristic of polar codes is that the channel orderings are channel-dependent. PO is based on the observation that the channel orderings are degraded according to the binary expansion of  their indexes. Therefore, in this method the channel orderings are not channel-dependent. PO provides us with quality information for all channels. This can be used to simplify code construction of polar codes. In \cite{Vangala} and \cite{Cheng} a comparative study of the performance of polar codes constructed by various techniques using the AWGN channel and SC decoder has been carried  out. Construction of polar codes based on the AWGN channel and GA have been reported in \cite{Wu} and \cite{Yuan}. In \cite{Vangala} the design of good polar codes with any construction method has been verified with the SC decoding error rate for various scenarios and polar code construction algorithms. Polar codes can also be constructed and adapted to a specific decoder, for example, construction of polar code for SCL decoding \cite{P. Yuan} and BP decoding \cite{M. Qin}\cite{S. Sun}. In \cite{A. Elkelesh} the authors propose a genetic algorithm framework that jointly optimizes the code construction and rate with a specific decoder.

To obtain code length flexibility, arbitrary kernels and multi-kernel (MK), puncturing, shortening and extension are typically performed. Polarization matrices of various sizes, for example 3x3, 5x5 and 7x7, were used to build a polar code of any length. BCH kernel matrices proposed in \cite{S.B. Korada} and the code decompositions proposed in \cite{N. Presman} both have restrictions on the size of the kernel matrices. Square polarizing kernels larger than two have been proposed in \cite{LZhang}, \cite{Serbetc} and \cite{Zhiliang}, while a polar code construction with mixed kernel sizes has been proposed in \cite{Gabry}\cite{Benammar}. By considering different polarizing kernels of alternate dimensions, MK improves block length flexibility. Although the general coding and decoding structure follows the same structure of Arikan's standard polar codes, there is an increased complexity with the generalization. Polar codes construction using the Reed-Muller (RM) rule \cite{Hussami}\cite{B. Li}\cite{M. Mondelli} can improve the performance of the error rate.

Various shortening and puncturing methods for polar codes have been reported in \cite{Bioglio}-\cite{MJang}. Generally, puncturing or shortening causes a loss of performance because when the number of bits punctured or shortened increases the code length decreases, degrading the performance. A review on puncturing and shortening techniques can be found in \cite{Bioglio}, which includes designs based on the column weights (CW) of the generator matrix and the reversal quasi-uniform puncturing scheme (RQUP) based on bit reversed permutation. In practice, the design of the code is made for $N = 2^n$ and rate-compatible code design techniques use a specifically chosen criterion for shortening or puncturing and generate a codeword with length $M$ that is $2^{n-1} < M < 2^n$. In \cite{Hong}-\cite{Chen} puncturing methods have been reported using belief propagation (BP) decoding based on optimization techniques employing retransmission  schemes such as Hybrid Automatic Repeat reQuest (HARQ). Different properties of punctured codes have been explored: minimum distance, exponent binding, stop tree drilling, and the reduced generating matrix method \cite{NiuChen}-\cite{EslamiPishro}. Schemes that depend on the analysis of density evolution were proposed in \cite{Kim} and \cite{Zhang}. On the other hand, shortening methods have been studied with SC decoding. In shortening techniques, we freeze a bit channel that receives a fixed zero value. The decoder, however, uses a plus infinity log-likelihood ratio (LLR) for that code bit as it is often assumed that this value is known. The study in \cite{Miloslavskaya} proposed a search algorithm to jointly optimize the shortening patterns and set of frozen bits. The work in \cite{Wang} devised a simple shortening method, reducing the generator matrix based on the weight of the columns (CW). In \cite{Niu} the reversal quasi-uniform puncturing scheme (RQUP) for reducing the generator matrix has been proposed. The authors in \cite{Oliveira} presented a polarization-driven (PD) shortening technique for the design of rate-compatible polar codes, which consists of reducing the generator matrix by relating its row index with the channel polarization. Recently, the PW algorithm has been used in a puncturing and shortening technique reported in \cite{MJang}.

Extension of polar codes have been studied in \cite{Saber}, \cite{MZhao} and \cite{Huang}. For HARQ schemes it was proposed that an arbitrary number of incremental coded bits can be generated by extending the polarization matrix such that multiple retransmissions are aggregated to produce a longer polar code with extra coding gain. The complexity of the encoder and decoder are similar to that of standard polar codes. Nevertheless, there is a significant increase in complexity when designing flexible-length polar codes by concatenated codes \cite{PTrifonov}, \cite{Mahdavifar} and by asymmetric kernel construction \cite{Cavatassi}. In both cases the code construction is specific to each kernel dimension without generalization gains. Chained polar subcodes \cite{Trifonov2} were shown to be an effective method for achieving length compatible polar codes.

In this paper, we present the concept of non-uniform polarization, which allows the construction of rate-compatible polar codes considering the polarization of channels with an arbitrary distribution of the Bhattacharyya parameter. Generalized channel polarization schemes have been proposed in \cite{Mahdavifar} and \cite{Kim2}. In \cite{Mahdavifar} the authors present a rate-matching scheme for multi-channel polar codes, which can be directly applied to bit-interleaved coded modulation schemes. Moreover, in \cite{Kim2} polar coding schemes are introduced for independent parallel channels in which the channel parameters are no longer fixed but exhibit some random behaviors.

The original construction of Arikan \cite{Arikan} considers a single Bhattacharyya parameter to all channels, which we call uniform polarization. We develop a Non-Uniform Polarization technique based on the Gaussian Approximation (NUPGA) for designing rate-compatible polar codes of arbitrary length \cite{nupga_access}. We then develop NUPGA-based shortening and extension algorithms for rate-compatible polar code designs. In the proposed techniques the shortened or extended channels are re-polarized for a more efficient choice of noiseless channels. We also present a generalization of the GA algorithm, which is used for both polarization of the initial channel and re-polarization of the shortened channel, and we present a simplified construct technique for extended polar codes. Simulations compare the proposed NUPGA techniques with existing approaches. A key feature of the proposed designs in that, the encoder and decoder structures are the same as that of the original polar codes \cite{Arikan} and require the same complexity.

In summary, the main contributions of this work are:
\begin{itemize}
  \item a novel NUPGA approach for designing rate-compatible polar codes along with a proof that it achieves capacity;
  \item NUPGA-based shortening and extension techniques for designing rate-compatible polar codes;
  \item an extensive simulation study that compares the NUPGA and existing design techniques.
\end{itemize}

The rest of the paper is organized as follows. In Section II, we present a brief review of the fundamentals of polar codes required for our exposition, outline their notation, and describe their encoding and SC/SCL decoding. In Section III, we review the principles of polarization theory of polar codes. In Section IV, we discuss the non-uniform polarization of polar codes and show that similarly to uniform polarization, it achieves channel capacity. In Section V, we give a detailed description of the proposed NUPGA-based shortening and extension techniques along with the pseudo-codes for their implementation. We present our simulation results in Section VI and the conclusions in Section VII.

\section{Polar codes}

In this section we review some fundamental aspects of polar codes that will help in the exposition of the proposed techniques. Using the notation proposed by Arikan \cite{Arikan}, we describe key principles of the encoder and the decoder of polar codes.

Given a B-DMC $W : \mathcal{X} \to \mathcal{Y}$, where $\mathcal{X} =$  $\{0,1\}$ and $\mathcal{Y}$ denote the input and output alphabets, respectively, we define the channel transition probabilities as $W(y|x)$, $x \in \mathcal{X}$, $y \in \mathcal{Y}$. As demonstrated in \cite{Arikan}, after channel combining and splitting operations on $N = 2^n$ independent uses of $W$, we get $N$ successive uses of synthesized binary input channels $W^{(i)}_N$ with $i = 1,2,\ldots,N$. $K$ represents the most reliable sub-channels. The indices of $K$ form the set of information $\mathcal{A}$. They are the ones that carry the information bits. The indices of the remaining sub-channels are included in the complementary set $\mathcal{A}^\text{c}$, and can be set to fixed bit values, all zeros.

A polar codes scheme is uniquely defined by three parameters: code-length $N = 2^n$, code-rate $R=K/N$ and an information set $\mathcal{A} \in$ $[N]$ with cardinality $K$. For encode we have $x^N_1= u^N_1\textbf{G}_N$, where $\textbf{G}_N$ is the transform matrix, $u^N_1 \in \{0,1\}^N$ is the source block and $x^N_1 \in \{0,1\}^N$ is the codeword. The source block $u^N_1$ is comprised of information bits $u_{\mathcal{A}}$ and frozen bits $u_{\mathcal{A}^\text{c}}$. The $N$-dimensional matrix can be recursively defined as $\textbf{G}_N = \textbf{B}_N\textbf{F}^{\otimes n}_2 $, where $\otimes$ denotes the Kronecker product, $\textbf{F}_2 =
\footnotesize\left[\begin{array}{cc}
1 & 0 \\
1 & 1 \end{array} \right]$ and $\textbf{B}_N$ is the bit-reversal permutation matrix, which can be omitted without loss of generality. Then, the codeword is transmitted to the receiver through AWGN channel.

We may adopt the SC decoder so that the information bits are estimated as \cite{Arikan}:

\begin{equation}
\hat{u}_i=\arg\max\limits_{u_i \in \{0,1\}} \ W_N^{(i)}(y_1^N,u_1^{i-1}|u_i),i \in A.
\end{equation}

Having known the decoded bits $\hat{u}^N_1=(\hat{u}_1,\ldots,\hat{u}_N)$ and received sequence $y^N_1=(y_1,\ldots,y_N)$, the likelihood ratio (LR) message of $u_i$, ${\rm LR}(u_i)=\frac{W^{(i)}_N(y^N_1,\hat{u}_1^{i-1}|0)}{W^{(i)}_N(y^N_1,\hat{u}_1^{i-1}|1)}$, can be recursively calculated using the SC decoding algorithm \cite{Arikan}. Then, $\hat{u_i}$, an estimated value of $u_i$, can be computed by:
\begin{equation}
  \hat{u}_i=\begin{cases}
    h_i(y_1^N,\hat{u}_1^{i-1}), & \text{if $i \in \mathcal{A}$}.\\
    u_i, & \text{if $i \in \mathcal{A}$$^c$}.
  \end{cases}
\end{equation}
where $h_i:\mathcal{Y}^\text{N} \times \mathcal{X}$ $^{i-1} \to \mathcal{X}$, $i \in \mathcal{A}$, are decision functions defined as:
\begin{equation}
  h_i(y_1^N,\hat{u}_1^{i-1})=\begin{cases}
    0, & \text{if $\frac{W^{(i)}_N(y^N_1,\hat{u}_1^{i-1}|0)}{W^{(i)}_N(y^N_1,\hat{u}_1^{i-1}|1)}\geq 1$}\\
    1, & \text{otherwise}.
  \end{cases}
\end{equation}
for $y_1^N \in \mathcal{Y}^\text{N}$, $\hat{u}_1$ $^{i-1} \in \mathcal{X}$ $^{i-1}$.

The LR output from the node at row $i$ and stage $j$ is denoted as $L(i,j)$, according to the decoding tree in \cite{Arikan}. In order to calculate $L(i,j)$, as proposed in \cite{Arikan}, we compute the following equation recursively:
\begin{equation}
  L(i,j+1)= \\
   \begin{cases}
    f(L(i,j),L(i+n/2^{j+1},j)), &\text{($f$ nodes)}\\
    g(L(i-n/2^{j+1},j),L(i,j),\hat{u}_{sum}), &\text{($g$ nodes)}
  \end{cases}
\end{equation}
where $f$ and $g$ functions were defined in \cite{Arikan} as:
\begin{equation}
f(a,b)=\frac{1+ab}{a+b}
\end{equation}
\begin{equation}
g(a,b,\hat{u}_{sum})=a^{1-2\hat{u}_{sum}}b
\end{equation}

Notice that in (6), $\hat{u}_{sum}$ is the modulo-2 sum of partial previous decoded bits. The term $\hat{u}_{sum}$ depicts the successive operation in the SC algorithm. The decision on the current bit strongly depends on the estimate of previously decoded bits. A log-likelihood ratio (LLR)-based SC algorithm has been developed in \cite{Leroux} to simplify the design. Accordingly, (5)-(6) in the natural domain are transformed to the following equations in the logarithm domain:
\begin{equation}
F(a,b)=2\text{tanh}^{-1}(\text{tanh}(a/2)\text{tanh}(b/2))
\end{equation}
\begin{equation}
G(a,b,\hat{u}_{sum})=a(-1)^{\hat{u}_{sum}}+b.
\end{equation}

Similar to LDPC decoding, a min-sum approximation \cite{Leroux} can be further employed to reduce the complexity of (7):
\begin{equation}
F(a,b) \approx \text{sign}(a)\text{sign}(b)\min(|a|,|b|),
\end{equation}
where (8)-(9) describe the LLR version of the  SC algorithm.

For the sucessive cancelation list (SCL) decoder \cite{Tal2}, let $S^{(i)}$ denote the set of candidate sequences in the $i$th step of the decoding process, and $|S^{(i)}|$ is the size
of $S^{(i)}$. Let $L$ be the maximum allowed size of the list and
$T$ be a threshold for pruning with $T \leq 1$. The SCL algorithm
can be described as follows:
\begin{itemize}
\item bits are estimated successively with index $i = 1, 2, \ldots, n$;
for each candidate in the list,

\item generate two $i$-length
sequences for decoding $\hat{u}_i$ as bit 0 and bit 1 by SC decoding;

\item if the number of candidates $|S^{(i)}|$ is not larger than $L$, there is nothing to do;

\item otherwise, reserve $L$ candidates with the largest probabilities and drop the others from $S^{(i)}$;

\item check each candidate $\hat{u}_i \in S^{(i)}$, if $P(\hat{u}_1^i)
< T \max\limits_{\hat{u}_1^i \in S^{(i)}}P(\hat{u}_1^i)$, eliminate
$\hat{u}_1^i$ from $S^{(i)}$.
\end{itemize}

For sequence estimation, after all the bits are examined,
re-encode every candidate in the list and calculate the corresponding
likelihood probabilities. Select the one with the maximal
probability as the sequence estimate:
\begin{equation}
\hat{u}_1^N=\arg\max\limits_{\hat{u}_1^N \in S^{(i)}} \prod\limits_{i=1}^{N} W(y_i|(u_1^N)_i).
\end{equation}

\section{Theoretical aspects of polar codes construction}

In this section, we recall the theoretical aspects of the construction of polar codes and show that the channel polarization theory can be generalized to non-uniform channels. In this scenario, the main aspects of the channel polarization theory are maintained, namely, the conservation of the associated channel capacity and the induction to the polarized channel \cite{Arikan}.

\subsection{PC construction}

To construct polar codes, it is necessary to find the set of indices for the information bits, $\mathcal{A}$. Standard polar codes can have length $N=2^n$ distinct bit-channels $(W_N)$. If $N \to \infty$, these bit-channels are divided into either noise free or completely noisy channels. To measure the quality of a binary input channel $W$, the Bhattacharyya parameter of $W$, which is indicated by $Z(W)$ \cite{Arikan} and defined as
\begin{equation}
Z(W) = \sum_{y \in \mathcal{Y}}\sqrt{W(y|x=0)W(y|x=1)},
\end{equation}
where $W(Y|X)$ is the channel transitional probability of input alphabet $\mathcal{X}$ $=\{0,1\}$ and output alphabet $\mathcal{Y}$, $x \in \mathcal{X}$, $y \in \mathcal{Y}$. For any binary discrete memoryless channel (BDMC), the reliability of bit-channels can be recursively determined \cite{Arikan}, and with the exception of the BEC channel, for all other channels its method of determination is approximate \cite{Arikan}, and as seen earlier, several algorithms have been proposed \cite{Vangala}. For example, BEC channels with $Z(W)$ close to zero are almost noiseless, whereas channels with $Z(W)$ close to one are almost pure-noise channels. The essential idea is to choose the most reliable bit-channels (noise free channels) to transmit information bits ($\mathcal{A}$), while noisy bit channels known to both encoder and decoder are frozen ($\mathcal{A}^\text{c}$).

For the construction of arbitrary-length polar codes, a generalization of channel polarization is necessary for the definition of non-uniform polarization, maintaining the primary results of the channel polarization theory. First, we will generalize channel types to see if full capacity is maintained. Then, we will verify if the channel polarization theory is valid for  non-uniform channels.

\subsection{Channel capacity}

Consider the discrete memoryless channel (DMC) shown in Fig. 1, where it is seen that the symbol $U$ is passed through the DMC designated by the symbol $W$.

\begin{figure}[htb]
\begin{center}
\includegraphics[scale=0.95]{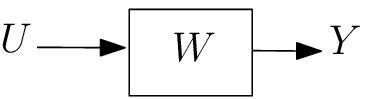}
\caption{Discrete memoryless channel with random variables}
\end{center}
\label{figura:fig_W}
\vspace{-1 em}
\end{figure}

The symbol at the input of the DMC can be considered as generated by a discrete random variable $U$. Similarly, the symbol at the output of the channel is modeled by another discrete random variable $Y$. Then, a set of DMC channels from Fig. 1 can be shown as in Fig. 2a. The Bhattacharyya parameter is $Z(W)$ for all DMC channels. Consider the transmission of $N$ different symbols $[u_1 u_2 \cdots u_N]$ through the channel in a serial manner. Consider that these $N$ symbols are generated by independent and identically distributed (i.i.d.) random variables. Without loss of generality, we consider that the transmission of each symbol is through each channel separately, as in Fig. 2a. Therefore, the deduction of the system's capacity will be the same as if we use other channels, that is, non-uniform channels, as suggested in Fig. 2b.

\begin{figure}[htb]
\begin{center}
\includegraphics[scale=0.95]{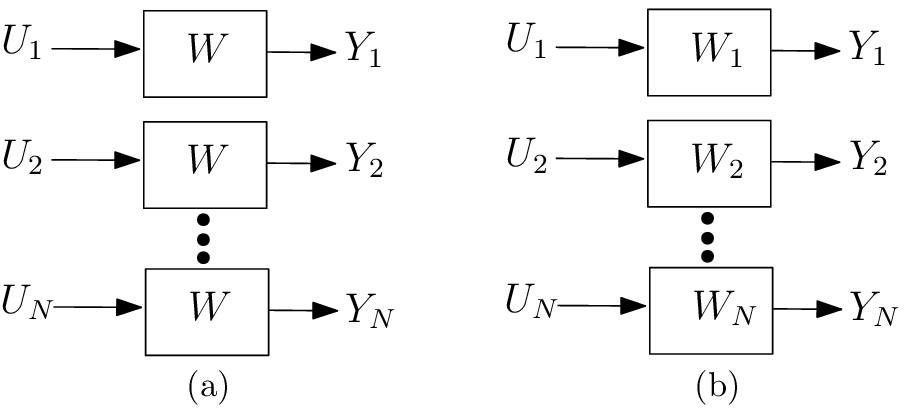}
\caption{(a) Uniform DMC and (b) non-uniform DMC }
\end{center}
\label{figura:fig_NW4}
\vspace{-1 em}
\end{figure}

Now, the Bhattacharyya parameter is different for all DMC channels. The mutual information for Fig. 2a and Fig. 2b are shown below.

Let $\textbf{U}=[U_1 \ U_2 \ \cdots \ U_N]$ and $\textbf{Y}=[Y_1 \ Y_2 \ \cdots \ Y_N]$ then the mutual information between $\textbf{U}$ and $\textbf{Y}$ can be written as
\begin{equation}
\begin{split}
I(\textbf{U};\textbf{Y}) & = I(U_1;\textbf{Y})+I(U_2;\textbf{Y}|U_1)+I(U_3;\textbf{Y}|U_1,U_2) \\
& \quad +\cdots +I(U_N;\textbf{Y}|U_1,U_2,\cdots,U_{N-1}). \label{eq:201}
\end{split}
\end{equation}
If $U_1$ is independent of $Y_1$, $Y_2$, $\cdots$, $Y_N$, then we can write that
\begin{equation}
\begin{split}
I(U_1;\textbf{Y}) & = I(U_1;Y_1) \\
I(U_2;\textbf{Y}|U_1) & = I(U_2;Y_2) \\
I(U_3;\textbf{Y}|U_1,U_2) & = I(U_3;Y_3) \\
I(U_N;\textbf{Y}|U_1,U_2,\cdots,U_{N-1}) & = I(U_N;Y_N) \nonumber
\end{split}
\end{equation}

Then, \eqref{eq:201} can be written as
\begin{equation}
I(\textbf{U};\textbf{Y}) = I(U_1;Y_1)+I(U_2;Y_2)+I(U_3;Y_3)+\cdots+I(U_N;Y_N)
\end{equation}
and let the capacity be $C = \text{max}I(\textbf{U};\textbf{Y})$, then we have
\begin{equation}
\begin{split}
\text{max}I(\textbf{U};\textbf{Y}) & = \text{max}I(U_1;Y_1)+\text{max}I(U_2;Y_2) \\
& \quad + \cdots +\text{max}I(U_N;Y_N) \\
\text{max}I(\textbf{U};\textbf{Y}) & = NC
\end{split}
\end{equation}
However, a key concept of channel polarization is that it consists of a method where the channel outputs depend on the other inputs as well. This implies the following inequalities:
\begin{equation}
\begin{split}
I(U_1;\textbf{Y}) & \neq I(U_1;Y_1) \\
I(U_2;\textbf{Y}|U_1) & \neq I(U_2;Y_2) \\
I(U_3;\textbf{Y}|U_1,U_2) & \neq I(U_3;Y_3) \\
I(U_N;\textbf{Y}|U_1,U_2,\cdots,U_{N-1}) & \neq I(U_N;Y_N) \nonumber
\end{split}
\end{equation}
That is, (12) cannot be simplified as in (13). In addition, it is necessary to ensure that
\begin{equation}
\begin{split}
I(U_1;\textbf{Y})& < I(U_3;\textbf{Y}|U_1,U_2) \leq I(U_2;\textbf{Y}|U_1)\\
< \cdots &< I(U_N;\textbf{Y}|U_1,U_2,\cdots,U_{N-1}),
\end{split}
\end{equation}
which means the individual capacities increase in an orderly manner but the total capacity remains constant, i.e., the total capacity of the channels is maintained, regardless of whether the channels are equal or not, that is, uniform or non-uniform, with different Bhattacharyya parameters. Then, for uniform channels, according to (14), and for non-uniform channels, with the inequality of (15), the capacity of the channels is conserved, and we show that non-uniform polarization schemes achieve symmetric capacity:
\begin{equation}
\text{max}I(\textbf{U};\textbf{Y}) = \sum_{i=1}^NC_i = NC
\end{equation}

Therefore, we can devise methods to construct polar codes that take into account different Bhattacharyya parameters. It is necessary to verify whether the theory of channel polarization \cite{Arikan} converges also for non-uniform channels. In the next sections, we show the main results for channel capacity, channel polarization and polarization convergence. Then, we show that all results remain valid for the case of generalized channels.

\subsection{Uniform construction}

In the classic construction of Arikan \cite{Arikan} the channel polarization considers $Z(W)$ identical for all channels. Let $W : \mathcal{X} \to \mathcal{Y}$ denote a general symmetric binary input discrete memoryless channel (B-DMC), with input alphabet $\mathcal{X}=$ $\{0,1\}$, output alphabet $\mathcal{Y}$ and the channel transition probability $W(y|x)$, $x \in \mathcal{X}$, $y \in \mathcal{Y}$. We write $W^N$ to denote the channel corresponding to $N$ uses of $W$: thus $W^N: \mathcal{X}$ $^N$ $\to \mathcal{Y}$ $^N$ with
\begin{equation}
W^N(y_1^N|x_1^N)=\prod_{i=1}^N W(y_i|x_i) \label{eq:1}
\end{equation}
The mutual information with equiprobable inputs, or symmetric capacity, is defined by \cite{Arikan}
\begin{equation}
I(W) = \sum\limits_{y \in Y}\sum\limits_{x \in X}\frac{1}{2}W(y|x)\log\frac{W(y|x)}{\frac{1}{2}W(y|0)+\frac{1}{2}W(y|1)},\label{eq:2}
\end{equation}
where the base-2 logarithm $0 \leq I(W) \leq 1$ is employed, and the corresponding reliability metric, the Bhattacharyya parameter is given by \cite{Arikan}
\begin{equation}
Z(W) = \sum\limits_{y \in Y}\sqrt{W(y|0)W(y|1)}, \label{eq:3}
\end{equation}
where $0 \leq Z(W) \leq 1$. For any B-DMC $W$, we have
\begin{equation}
log\frac{2}{1+Z(W)} \leq I(W) \leq \sqrt{1+Z(W)^2}. \label{eq:4}
\end{equation}

Applying the channel polarization transform for $N$ independent uses of $W$, after channel combining and splitting operation we obtain the group of polarized channels $W_N^{(i)}: \mathcal{X} \to \mathcal{Y} \times \mathcal{X}^{\text{i-1}}$, $i=1,2, \ldots ,N$, defined by the transition probabilities
\begin{equation}
W_N^{(i)}(y_1^N,u_1^{(i-1)}|u_i) = \sum\limits_{u_{i+1}^N \in X^{N-1}}\frac{1}{2^{N-1}}W_N(y_1^N|u_1^N), \label{eq:5}
\end{equation}
where $N=2^n$ is the code length.

Channel polarization is defined as an operation by which one generates $N$ independent copies of a given B-DMC $W$ and a second set of $N$ channels {$W_N^{(i)}: 1 \leq i \leq N$} that show a polarization effect in the sense that, as $N$ becomes large, the symmetric capacity terms {$I(W_N^{(i)})$} tend towards $0$ or $1$ for all but a vanishing fraction of indices $i$.

\begin{figure}[htb]
\begin{center}
\includegraphics[scale=0.95]{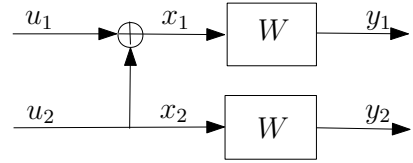}
\caption{The Channel $W_2$}
\end{center}
\label{figura:fig01}
\vspace{-1 em}
\end{figure}

The recursion combines two independent copies of $W$, as shown in Fig. 3, and obtains the channel $W_2: \mathcal{X}^\text{2} \to \mathcal{Y}^\text{2}$ with the transition probabilities
\begin{equation}
W_2^{(1)}(y_1^2|u_1)=\sum_{u_2}\frac{1}{2}W(y_1|u_1 \oplus u_2)W(y_2|u_2). \label{eq:6}
\end{equation}
\begin{equation}
W_2^{(2)}(y_1^2|u_1|u_2)=\frac{1}{2}W(y_1|u_1 \oplus u_2)W(y_2|u_2). \label{eq:7}
\end{equation}

The construction tree used in channel polarization is shown in Fig. 4 \cite{Arikan}. Note that it considers a uniform value of the Bhattacharyya parameter that is represented in the tree by $W$.

\begin{figure}[htb]
\begin{center}
\includegraphics[scale=0.95]{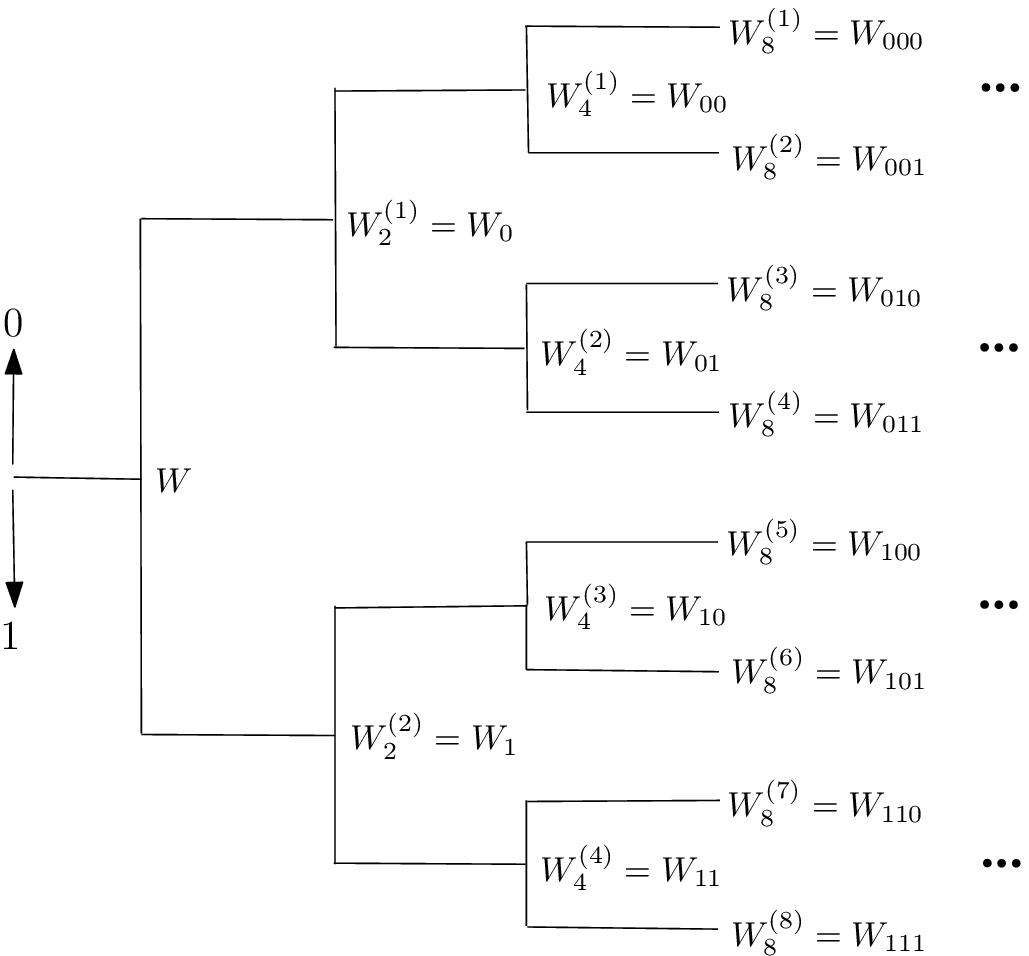}
\caption{The tree process for the recursive channel construction.}
\end{center}
\label{figura:tree_uniform}
\vspace{-1 em}
\end{figure}

For example, we show a set of results for BEC channels. Consider the transformation $(W_N^{(i)},W_N^{(i)}) \to (W_{2N}^{(2i-1)},W_{2N}^{(2i)})$ is rate-preserving and realiability-improving in the sense that \cite{Arikan}, and produces as output the following two binary-input DMCs (with larger output alphabets):
\begin{equation}
\begin{split}
(W_N^{(i)},W_N^{(i)}) & \to (W_{2N}^{(2i-1)},W_{2N}^{(2i)}) \\
Z(W_{2N}^{(2i)})  & \leq 2Z(W_N^{(i)})-Z(W_N^{(i)})^2 \\
 Z(W_{2N}^{(2i-1)}) & \leq Z(W_N^{(i)})^2 \\
 Z(W_{2N}^{(2i-1)}) & \leq Z(W_{2N}^{(2i)}) \label{eq:8}
\end{split}
\end{equation}
and that by definition in (18), we have
\begin{equation}
\begin{split}
I(W_{2N}^{(2i)}) & = I(U_1;Y_1,Y_2),      \\
I(W_{2N}^{(2i-1)}) & = I(U_2;Y_1,Y_2,U_1), \label{eq:9}
\end{split}
\end{equation}
where $U_1$ and $U_2$ are iid. From the chain rule, it follows that
\begin{equation}
\begin{split}
I(W_{2N}^{(2i)}) + I(W_{2N}^{(2i-1)}) & = I(U_1;Y_1,Y_2) + I(U_2;Y_1,Y_2,U_1) \\
 & = 2I(W_N^{(i)}) \label{eq:10}
\end{split}
\end{equation}
and
\begin{equation}
\begin{split}
I(W_{2N}^{(2i)}) & = I(U_2;Y_1,Y_2,U_1) \\
 & \geq I(W_N^{(i)}), \label{eq:11}
\end{split}
\end{equation}
which results in
\begin{equation}
I(W_{2N}^{(2i)}) \geq I(W_{2N}^{(2i-1)}) \label{eq:12}
\end{equation}

For the case of the parameter $Z$ in the BEC channel we have that the reliability terms further satisfy \cite{Arikan}
\begin{equation}
Z(W_{2N}^{(2i-1)}) \leq 2Z(W_N^{(i)})-Z(W_N^{(i)})^2 \label{eq:18}
\end{equation}
\begin{equation}
Z(W_{2N}^{(2i)}) = Z(W_N^{(i)})^2. \label{eq:19}
\end{equation}
The cumulative rate and reliability satisfy \cite{Arikan}
\begin{equation}
\sum_{i=1}^N I(W_{N}^{(i)}) = NI(W). \label{eq:20}
\end{equation}
\begin{equation}
\sum_{i=1}^N Z(W_{N}^{(i)}) \leq NZ(W). \label{eq:21}
\end{equation}

The uniform channel polarization for any set of B-DMCs $W_{(i)}$, $i \in [1,N]$, ensures that for arbitrary small $\delta \leq 0$ there exist polar codes that achieve the sum capacity $I_s$, in the sense that as $N\rightarrow \infty$, which is a power of 2, the fraction of indices $i \in [1,N]$ satisfies
\begin{equation}
\begin{split}
\frac{|\{i|I(W_N^{(i)}) \in (1-\delta,1]\}|}{N} \to \frac{I_s}{N}, \\
\frac{|\{i|I(W_N^{(i)}) \in [0,\delta)\}|}{N} \to 1-\frac{I_s}{N}, \label{eq:22}
\end{split}
\end{equation}
where the values of $I_N^{(i)}$ converge to $\{0,1\}$.

\section{Proposed Non-uniform Construction}

In this section, we show a generalization of the equations presented in Section III.B, and we verify that for non-identical channels, that is, with different Bhattacharyya parameters, they also converge for channel polarization.

If the channels are independent but not uniform, then $W_{(i)}:
\mathcal{X}_{\text{(i)}} \to \mathcal{Y}_{\text{(i)}}$, as shown in Fig. 5, where we rewrite \eqref{eq:1} as
\begin{equation}
W^{N'}(y_1^{N'}|x_1^{N'})=\prod_{i=1}^{N'} W_{(i)}(y_i|x_i).
\end{equation}
\begin{figure}[htb]
\begin{center}
\includegraphics[scale=0.95]{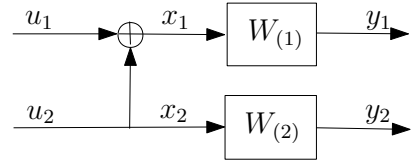}
\caption{The non-uniform channel $W_{2'}$ }
\end{center}
\label{figura:fig02}
\vspace{-1 em}
\end{figure}

We consider $N$ and $N'$ with the same cardinality and in such a way
that their transition probabilities $W_{(i)}(y|x)$ and $W_{(j)}(y|x)$ may differ if $i \neq j$. A non-identical channel corresponds to a channel with different Bhattacharyya parameters. As in \cite{Arikan}, given any B-DMC $W_{(i)}$ the same definitions of the symmetric capacity \eqref{eq:2} and the Bhattacharyya parameter \eqref{eq:3} are adopted as performance measures, and rewritten as
\begin{equation}
I(W_{(i)}) = \sum\limits_{y \in Y}\sum\limits_{x \in X}\frac{1}{2}W_{(i)}(y|x)\log\frac{W_{(i)}(y|x)}{\frac{1}{2}W_{(i)}(y|0)+\frac{1}{2}W_{(i)}(y|1)},
\end{equation}
\begin{equation}
Z(W_{(i)}) = \sum\limits_{y_i \in Y}\sqrt{W_{(i)}(y_i|0)W_{(i)}(y_i|1)}.
\end{equation}
We notice that the relation in \eqref{eq:4} is equivalent and can be
rewritten as
\begin{equation}
\log\frac{2}{1+Z(W_{(i)})} \leq I(W_{(i)}) \leq \sqrt{1+Z(W_{(i)})^2},
\end{equation}
where \eqref{eq:5} remains valid.

The $W_2$ channel with the transition probabilities of \eqref{eq:6} and \eqref{eq:7} can be written as
\begin{equation}
W_2^{(1)}(y_1^2|u_1)=\sum_{u_2}\frac{1}{2}W_{(1)}(y_1|u_1 \oplus u_2)W_{(2)}(y_2|u_2),
\end{equation}
\begin{equation}
W_2^{(2)}(y_1^2,u_1|u_2)=\frac{1}{2}W_{(1)}(y_1|u_1 \oplus u_2)W_{(2)}(y_2|u_2).
\end{equation}

Using the BEC channel again as an example, for a comparison with Section III.B, for the case of the parameter $Z$ we have:
\begin{equation}
\begin{split}
Z(W_2^{(2)}) & = \sum_{y_1^2,u_1}\sqrt{W_2^{(2)}(y_1^2,u_1|u_2=0)W_2^{(2)}(y_1^2,u_1|u_2=1)} \\
 & = \sum_{y_1^2,u_1}\frac{1}{2}\sqrt{W_{(1)}(y_1|u_1)W_{(2)}(y_2|0)} \\
 & \quad \cdot \sqrt{W_{(1)}(y_1|u_1)W_{(2)}(y_2|1)} \\
 & = \sum_{y_2,u_1}\sqrt{W_{(2)}(y_2|0)W_{(2)}(y_2|1)} \\
 & \quad \cdot \sum_{y_1,u_1}\frac{1}{2}\sqrt{W_{(1)}(y_1|u_1)W_{(1)}(y_1|u_1)} \\
 & = Z(W_{(2)})Z(W_{(1)}).
\end{split}
\end{equation}
The reliability terms in \eqref{eq:18} and in \eqref{eq:19} are rewritten as
\begin{equation}
Z(W_{2}^{(1)}) \leq Z(W_{(1)})+Z(W_{(2)})-Z(W_{(1)})Z(W_{(2)}),
\end{equation}
\begin{equation}
Z(W_{2}^{(2)}) \leq Z(W_{(1)})Z(W_{(2)}),
\end{equation}
\begin{equation}
Z(W_{2}^{(2)}) \leq Z(W_{2}^{(1)}).
\end{equation}
The cumulative rate in \eqref{eq:20} and the reliability in \eqref{eq:21} are then given by
\begin{equation}
\sum_{i=1}^{N'} I(W_{N'}^{(i)}) = \sum_{i=1}^{N'} I(W_{(i)}),
\end{equation}
\begin{equation}
\sum_{i=1}^{N'} Z(W_{N'}^{(i)}) \leq \sum_{i=1}^{N'} Z(W_{(i)}).
\end{equation}

\begin{figure}[htb]
\begin{center}
\includegraphics[scale=0.5]{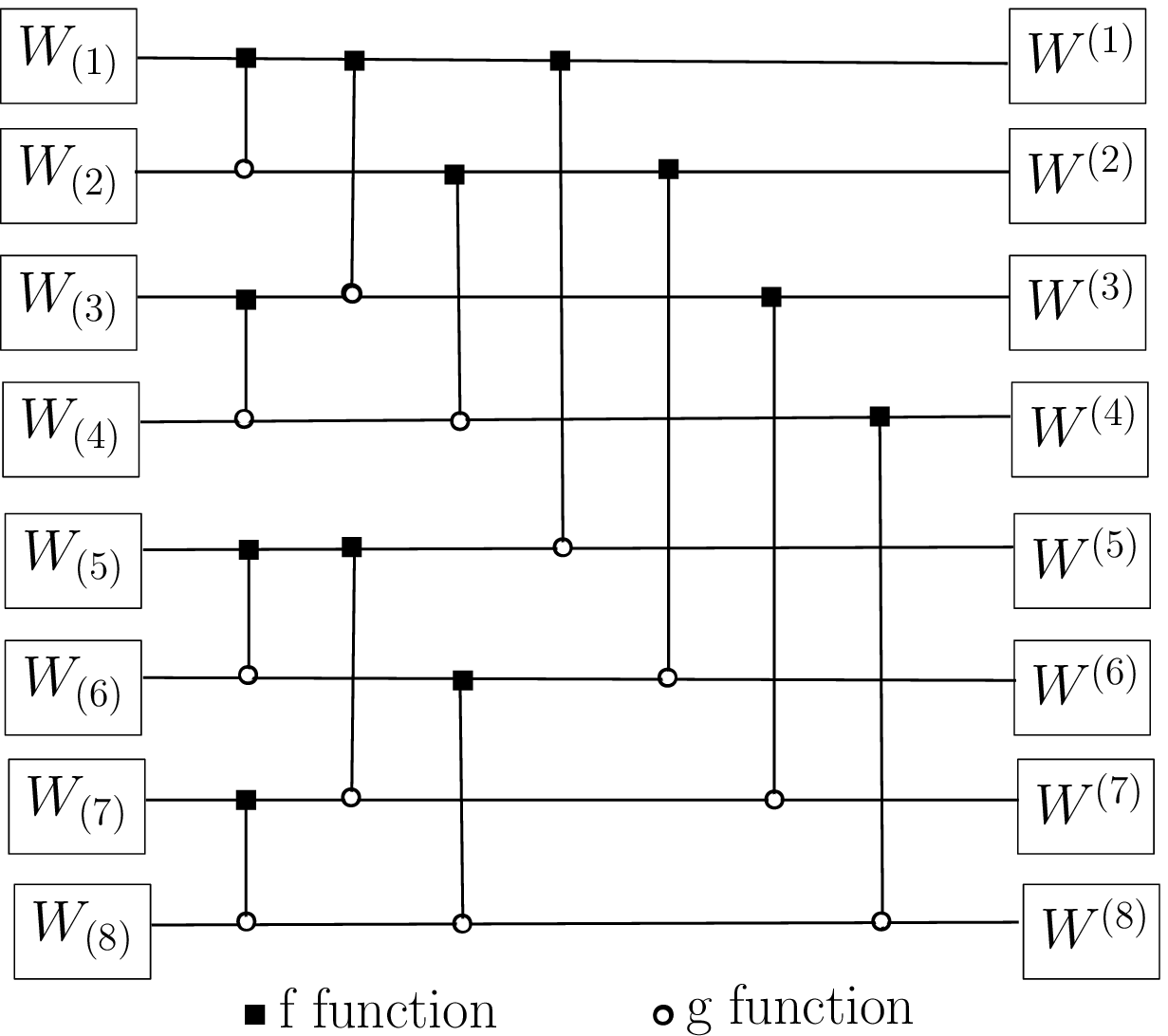}
\caption{The full polarization tree alternative.}
\end{center}
\label{figura:fig}
\vspace{-1 em}
\end{figure}

We can show that \eqref{eq:10} can be obtained by performing the following operations:
\begin{equation}
\begin{split}
I(W_2^{(1)}) & = I(Y_1,Y_2,U_1), \\
I(W_2^{(2)}) & = I(Y_1,Y_2,U_1;U_2), \\
I(W_2^{(1)}) + I(W_2^{(2)}) & = I(Y_1,Y_2,U_1) + I(Y_1,Y_2,U_1;U_2) \\
 & = I(W_1) + I(W_2). \nonumber
\end{split}
\end{equation}
The non-uniform polarization channel for any set of B-DMCs $W_N^{(i)}$, $i \in [1,N]$, for arbitrary small $\delta \leq 0$ there exist polar codes that achieve the sum capacity $I_s=\sum_{1}^{N}I_i$, $I_i=I(X_i,Y_i)$ in the sense that as $N\rightarrow \infty$, which is a power of 2, the fraction of indices $i \in [1,N]$ satisfies
\begin{equation}
\begin{split}
\lim\limits_{i \to \infty}\frac{|\{i|\sum_{1}^{N}I(W_N^{(i)}) \in (1-\delta,1]\}|}{2^i} \to \frac{I_s}{N}, \\
\lim\limits_{i \to \infty}\frac{|\{i|\sum_{1}^{N}I(W_N^{(i)})) \in [0,\delta)\}|}{2^i} \to 1-\frac{I_s}{N}.
\end{split}
\end{equation}
where the values of $I_N^{(i)}$ converge to $\{0,1\}$, which is a novel result related to the established result for uniform channel polarization in \eqref{eq:22}.

\section{Proposed NUPGA Design Algorithms}

In this section, we describe the proposed NUPGA design algorithms which employ the non-uniform construction. In the proposed NUPGA design algorithms, the tree process for the recursive channel polarization can be redesigned according to Fig.6 without loss of generalization, with the nodes being calculated using the functions described in (29) and (30).

\subsection{Proposed NUPGA-based shortening}

We use the PD shortening technique \cite{Oliveira} as a starting point  to define the channels that will initially be shortened. We recall that, the purpose of a shortening technique is to reduce the codeword length $N$ to $M$, that is $2^{n- 1} < M < 2^n$. Let $K$ denote the number of information bits. Let $N$  denote the code length of the basic polar codes, and let $M$ denote the code length of the shortened polar codes, where $K < M < N$. Let $P$ denote the shortening pattern, which is the index set of the shortened bits, and $|P| = N-M$ denote the cardinality of shortened bits. The code rate of the shortened codes is $R = K/M$. In shortening, the $P$ shortened bits of $x^N_1$ are known at the decoder, such that the LLRs of them can be set to infinity for decoding. Consider that the vector $P$ contains the channels obtained by the PD shortening technique \cite{Oliveira}. We first generate the codeword by defining the channel in $P$ as zero valued (frozen bits).  We then reduce the length of the codeword message based on the $P$. Polar codes are characterized by non-universality \cite{Arikan}. In this work, it is adopted the design-SNR equal to zero. As the code has been shortened, the reliability of the bit channels changes and the information set should change accordingly. In this regard, the study in \cite{Oliveira} indicates that the order of channel polarization does not change after shortening. For the shortened channels, we consider the parameter $Z(W)$ penalized a frozen bit, and will be used as input in the NUPGA method.

Under AWGN channels, the LLRs of each subchannel, namely $L_N^{(i)}$, the estimation of the channel polarization are calculated using the GA \cite{Trifonov} algorithm with the following recursions:
\begin{equation}
\begin{cases}
E(L^{(2i-1)}_N) = \phi^{-1}(1-(1-\phi(E(L^{(i)}_{N/2})))^2)\\
E(L^{(2i)}_N) = 2E(L^{(i)}_{N/2}), \label{eq:23}
\end{cases}
\end{equation}
where $E[\cdot]$ represents the mean of random variable, and
\begin{equation}
\phi(x) =
\begin{cases}
\exp(-0.4527x^{(0.86)}+0.0218) \ {\rm if} \ 0 < x \leq 10\\
\sqrt{\frac{\pi}{x}}(1-\frac{10}{7x})\exp(-\frac{x}{4}) \ ~~~~~~~~~~~~~ {\rm if} \ x > 10\\
\end{cases}
\end{equation}
With the generalization we have modified \eqref{eq:23} in order to make it applicable to arbitrary code lengths, with $f=E(L^{(2i-1)}_{N'})$ and $g=E(L^{(2i)}_{N'})$ according to Fig.6. This results in the following proposed recursions:
\begin{equation}
\begin{cases}
E(L^{(2i-1)}_{N'}) = \phi^{-1}(1-(1-\phi(E(L^{(i)}_1)))(1-\phi(E(L^{(i)}_2))))\\
E(L^{(2i)}_{N'}) = E(L^{(i)}_1)E(L^{(i)}_2).
\end{cases}
\end{equation}
The proposed NUPGA shortening algorithm is described in Algorithm 1. As an example,  consider $N = 4$ and $K = 2$ and initially the scheme without shortening. Then we have that the non-shortened vector is $P = \{1,1,1,1\}$, which is the expected result for $F = \{0,1,0,1\}$. Now consider the shortened vector $P = \{1,1,1,0\}$, which is applied in the operation indicated in step 5 of Algorithm 1. Then, the evolved bit channel capabilities are $Z([W^{(i)}_N])=\{0.21,1.64,2.28,0\}$ and the new information set $F = \{0,1,1,0\}$.

\begin{algorithm}
\caption{Proposed NUPGA Shortening Algorithm}
\begin{algorithmic}[1]
\State $\textbf{INPUT}$: $N,K,P$ and design-SNR $E_{dB}=(RE_b/N_o)$ in dB
\State $\textbf{OUTPUT}$: $F \in \{0,1,\ldots,N-1\}$ with
$|F|=N$
\State $S=10^{EdB/10}$ and $n=log_2N$
\State $L \in R^N$, Initialize $[E(L^{(i)}_1]^N_1=4S$
\State Upgrade with shortening vector $[E(L^{(i)}_1)]^N_1$ with $P$
\For{i = 1 to $n+1$}
\State $d=2^{(i-2)}$
\For{b = 1 do $2^{(i-1)}$ to $N$}
\For{k = 0 to $d-1$}
\If {$E(L^{(i-1)}_{k+b})=0$ or $E(L^{(i-1)}_{k+b+d})=0$}
\State $E(L^{(i)}_{k+b})=E(L^{(i-1)}_{k+b})$
\State $E(L^{(i)}_{k+b+d})=E(L^{(i-1)}_{k+b+d})$
\EndIf $\textbf{end if}$
\State $E(L^{(i)}_{k+b}) = \phi^{-1}(1-(1-\phi(E(L^{(i-1)}_{k+b})))(1-$
\State $\ \ \ \ \ \ \ \ \ \ \ \ \ \ \ \ \ \ \ \ \phi(E(L^{(i-1)}_{k+b+d}))))$
\State $E(L^{(i)}_{k+b+d}) = E(L^{(i-1)}_{k+b})E(L^{(i-1)}_{k+b+d})$
\EndFor $\textbf{end for}$
\EndFor $\textbf{end for}$
\EndFor $\textbf{end for}$
\State $F={\rm Find ~indices ~of ~ smallest ~elements~}(E[L],K)$
\State return $F$
\end{algorithmic}
\end{algorithm}

\subsection{Proposed NUPGA-based extension}

A simple polar codes extension scheme can be implemented as suggested in Fig.7 \cite{Huang}. An additional level of polarization is performed in $P_1^M$ and the information bits $u_1^k$. The connection between the additional channels $P_1^M$ and channels $u_{k-M}^k$ is carried out by the linear polarization sequence of $u_1^k$. In the extension scheme the complexity is $NlogN$, less than the puncturing/shortening scheme which is $(N+1)log(N+1)$. This scheme is efficient for kernels with low dimension ($N<512$) and for extension of $P<50\%$ of $N$. Using the proposed NUPGA technique, we can consider all additional bit channels as $P_1^M=0$ and as bits output from polarized channels $u_{k-M}^k$, uniform and limited to the length of the extension. For the encoder we have the same definition, that is, $P_1^M=0$. In the decoder, we have
\begin{equation}
\hat{u}_1^k= f(LLR((u_1^k)+LLR(u_{k-M}^k)),
\end{equation}
in the same order.
\begin{figure}[htb]
\begin{center}
\includegraphics[scale=0.95]{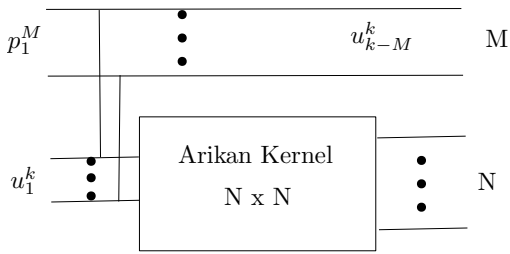}
\caption{Polar code extension scheme}
\end{center}
\label{figura:extension}
\vspace{-1 em}
\end{figure}

Note that according to (44), we have
\begin{equation}
\sum_{i=1}^{N} I(W_{N}^{(i)})+\sum_{i=1}^{M} I(W_{M}^{(i)}) = \sum_{i=1}^{N} I(W_{(i)})+\sum_{i=1}^{M} I(W_{(i)}) \\ \nonumber
\end{equation}
this only happens in the following two cases: either $W_N^{(i)}$ and
$W_M^{(i)}$ are noise channel and both of the $I(\cdot)$  are equal
to 1, or $W_N^{(i)}$ and $W_M^{(i)}$ are both perfect channel such
that the two $I(\cdot)$  are equal to 0. Using NUPGA, when
$W_N^{(i)}$ is perfect and $W_M^{(i)}$ is useless, the two
$I(\cdot)$ are 1 and 0, respectively. In other words, if the
extended bit channel $W_M^{(i)}$ is noise channel and excluding the
case that both $W_N^{(i)}$ and $W_M^{(i)}$ are perfect channel, the
re-polarization improves the reliability of the shortened channels.
With regards to (45), we have
\begin{equation}
\sum_{i=1}^{N} Z(W_{N}^{(i)})+\sum_{i=1}^{M} Z(W_{M}^{(i)}) \leq \sum_{i=1}^{N} Z(W_{(i)})+\sum_{i=1}^{M} Z(W_{(i)}) \nonumber
\end{equation}
and with the use of NUPGA, we have
\begin{equation}
\sum_{i=1}^{M} Z(W_{M}^{(i)}) \leq \sum_{i=1}^{N} Z(W_{N}^{(i)}) \nonumber
\end{equation}
and then
\begin{equation}
\sum_{i=1}^{M} Z(W_{(i)}) \leq \sum_{i=1}^{N} Z(W_{(i)}) \nonumber
\end{equation}
which ensures that the extended channels will all be noisy.

This method is similar to the extension of the polarization matrix
proposed in \cite{MZhao} and \cite{Huang}. Note that the
construction method for polar codes extension allows us to maintain
the same encoder and decoder for codeword N. In the proposed NUPGA
Extension algorithm, the original codeword of the channels initially
designed is first extended by adding new bits. Thus, it is possible
to increase the code length by gradually adding new bits, making it
possible to build codewords of any length. The information bit
channels are polarized according to the reliability of the bit
channel calculated from the new extended channels. The main idea of
the proposed NUPGA extension algorithm is to generate the new bit
channels as frozen bits and make the associated information bits
more reliable than before. The extension length is $\Delta M$ and
the new rate is $M = (N + \Delta M)$ which can still be decoded
efficiently. Therefore, the proposed NUPGA extension algorithm is
employed for the design of extension channels. The details of the
proposed NUPGA extension algorithm are shown in Algorithm 2.

\begin{algorithm}
\caption{Proposed NUPGA Extension Algorithm}
\begin{algorithmic}[1]
\State $\textbf{INPUT}$: $N,K,P,\Delta M$ and design-SNR $E_{dB}=(RE_b/N_o)$ in dB
\State $\textbf{OUTPUT}$: $F \in \{0,1,\ldots,N+\Delta M-1\}$
\State $S=10^{EdB/10}$, $n=log_2N$
\State $L \in R^{N}$,
Initialize $[E(L^{(i)}_1)]^N_1=4S$ and $[E(L^{(i)}_1)]^{\Delta M}_N=0$
\State Upgrade with vector $[E(L^{(i)}_1)]^N_1$ with $P$
\For{i = 1 to $n+1$}
\State $d=2^{(i-2)}$
\For{b = 1 do $2^{(i-1)}$ to $N$}
\For{k = 0 to $d-1$}
\If {$E(L^{(i-1)}_{k+b})=0$ or $E(L^{(i-1)}_{k+b+d})=0$}
\State $E(L^{(i)}_{k+b})=E(L^{(i-1)}_{k+b})$
\State $E(L^{(i)}_{k+b+d})=E(L^{(i-1)}_{k+b+d})$
\EndIf $\textbf{end if}$
\State $E(L^{(i)}_{k+b}) = \phi^{-1}(1-(1-\phi(E(L^{(i-1)}_{k+b})))(1-$
\State $\ \ \ \ \ \ \ \ \ \ \ \ \ \ \ \ \ \ \ \ \phi(E(L^{(i-1)}_{k+b+d}))))$
\State $E(L^{(i)}_{k+b+d}) = E(L^{(i-1)}_{k+b})E(L^{(i-1)}_{k+b+d})$
\EndFor $\textbf{end for}$
\EndFor $\textbf{end for}$
\EndFor $\textbf{end for}$
\State $F={\rm Find ~indices ~of ~ smallest ~elements~}(E[L],K)$
\State return $F$
\end{algorithmic}
\end{algorithm}

\section{Simulation Results}

In this section, we evaluate the performance of the proposed NUPGA algorithms described in Section V and compare them with the shortened polar code techniques CW \cite{Wang}, RQUP \cite{Niu} and PD \cite{Oliveira}. We assess the Bit Error Rate (BER) and Frame Error Rate (FER) performances of the polar codes using BPSK for data transmission over an AWGN channel. We consider different shortened codewords, rates and variations of the SC decoder and the SC list (SCL) decoder. In Fig.8 we show the performance of the proposed NUPGA shortening and existing algorithms under the SC decoder as described in \cite{Vangala}, whereas Fig.9 shows their performance under SCL \cite{Tal2} with list size $L=16$. Fig.10 shows the performance of the proposed NUPGA shortening and existing techniques using a Cyclic Redundancy Check (CRC)-aided list decoding (CA-SLC) \cite{Tal2} with list size $L=16$ and CRC with size 24. In Fig.11 we compare the performance of the proposed NUPGA extension algorithm with the PD and NUPGA shortening algorithms with the list decoder.

\begin{figure}[htb]
\begin{center}
\includegraphics[scale=0.45]{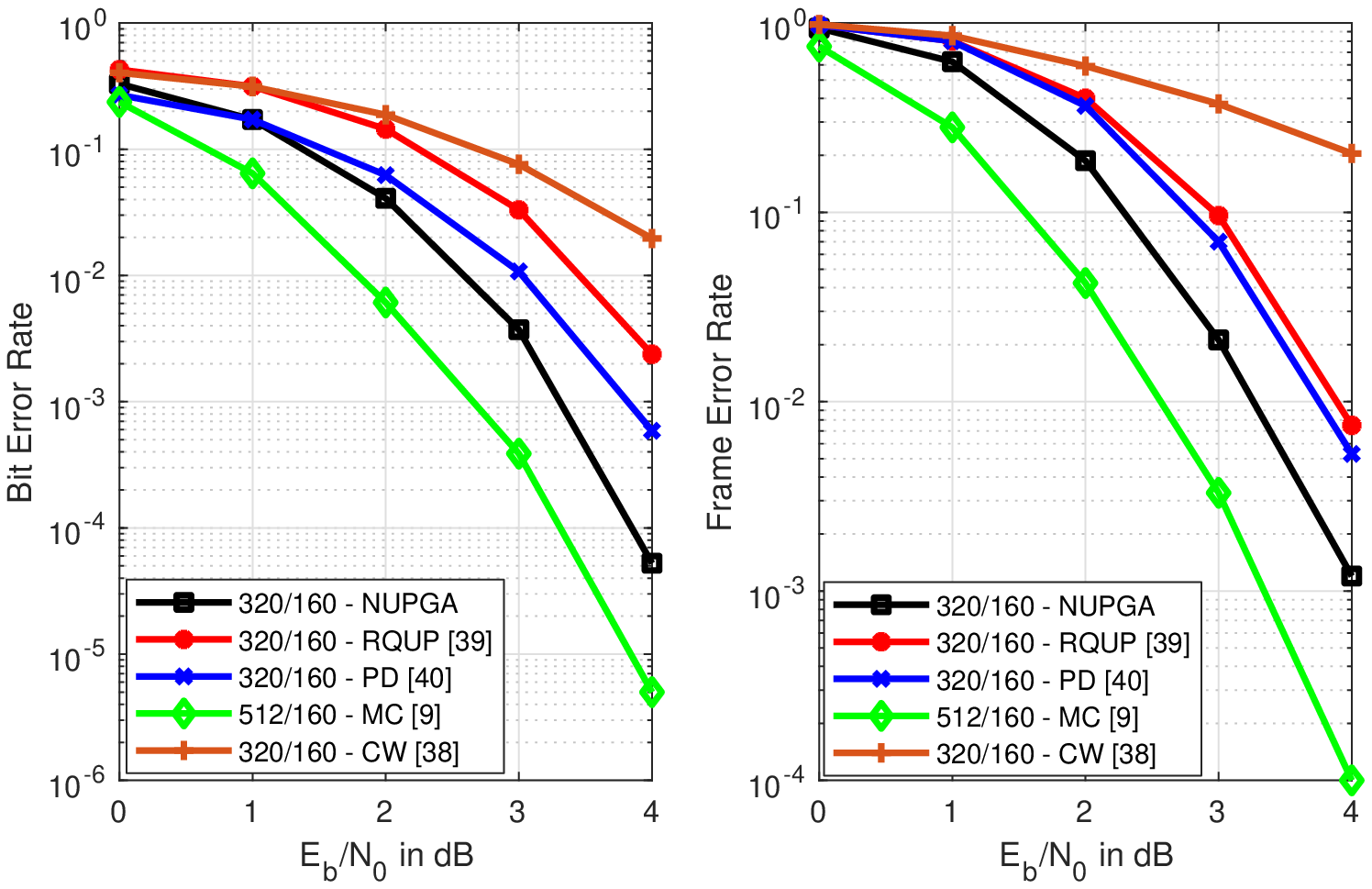}
\vspace{-0.5 em}
\caption{Performance of polar codes with $N=512$ and rate-compatible
codewords using shortening with $M=320$ and $K=160$.}
\end{center}
\label{figura:fig_ISWCS_512_god}
\vspace{-0.5 em}
\end{figure}

In the first example, in Fig.8, we show the performance of Arikan's
standard polar codes \cite{Vangala} with length $N=512$, denoted as
mother code (MC), and their rate-compatible versions operating at
rate $R = 1/2$ with $M =320$ and $K = 160$ using the CW \cite{Wang},
RQUP \cite{Niu}, PD \cite{Oliveira} and the proposed NUPGA
shortening techniques. In Fig.9 we show the performance of codes of
length $M = 400$, $K = 200$ for the rate-compatible designs and
CA-SCL with $L=16$. In Fig.10 we show the performance of codes of
length $M = 400$, $K = 50$, CA-SCL with $L=16$ and CRC with length
24 in a system operating with rate $R=1/4$. In Fig.11, we compare
the performance of the shortened polar codes $M = 280$ and $k =
128$, the proposed NUPGA extension technique with three other
curves: with MC \cite{Vangala} of length $N = 256$, the PD
\cite{Oliveira} algorithm and the  proposed NUPGA shortening
algorithm under list decoding (CA-SCL) with $L = 16$ and CRC with
length 24. The results show that the NUPAG extension technique
outperforms the NUPGA shortening and the PD algorithms using list
decoding with CRC.

\begin{figure}[htb]
\begin{center}
\includegraphics[scale=0.45]{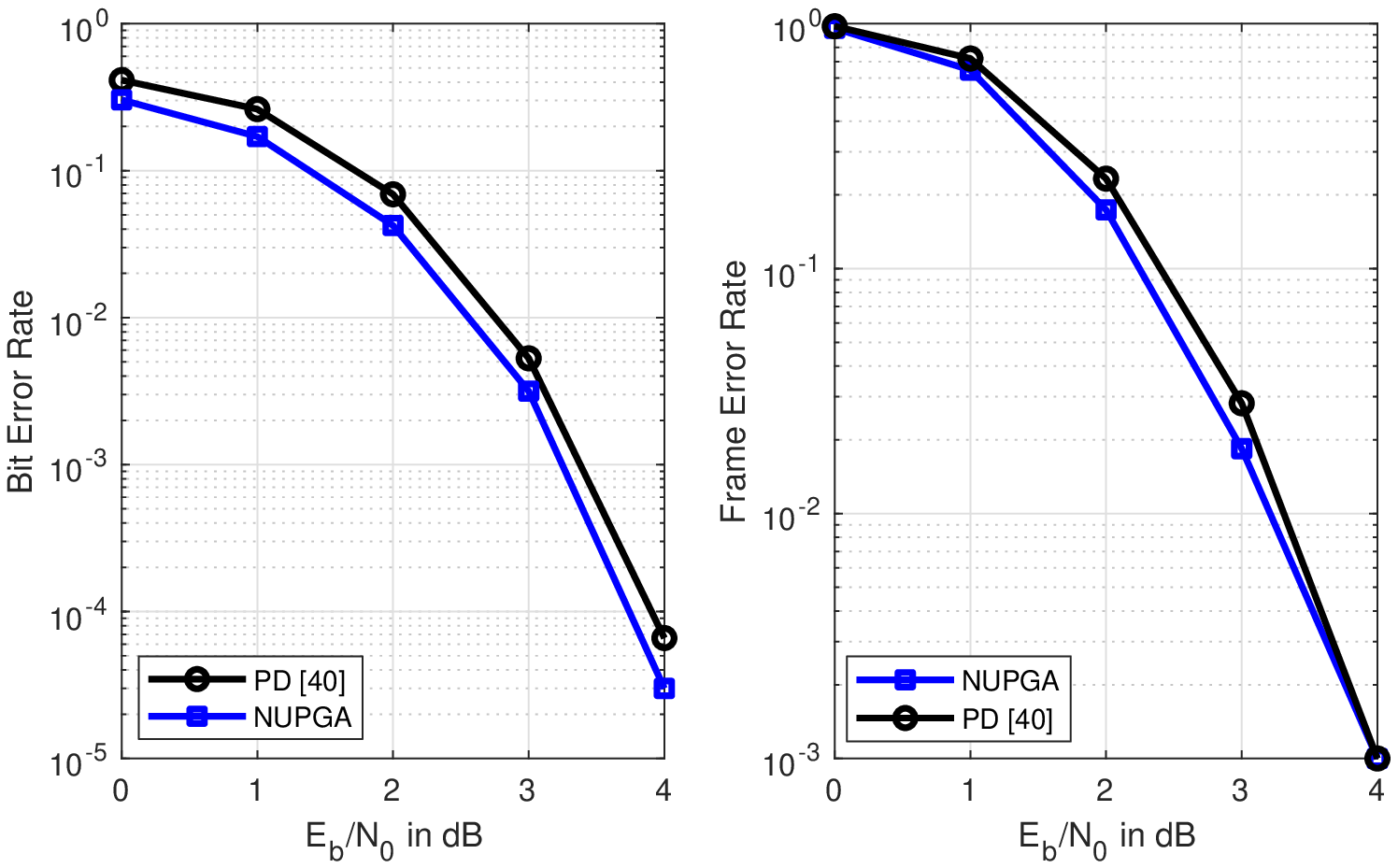}
\vspace{-0.5 em}
\caption{Performance of PD and NUPGA shortening algorithm, both with $M=400$, $K=200$ and CA-SCL with $L=16$.}
\end{center}
\label{figura:fig_400_200_L16}
\vspace{-0.5 em}
\end{figure}

We can observe a performance gain of the NUPGA shortening algorithm in all simulations, and the gain is greater in the case of low rates under list decoding, as shown in Fig.10. In Fig.8 we observe that the performance gain of NUPGA is of the order of 0.5 dB as compared to the approach of CW \cite{Wang}. In Fig.9, we observe that the gain obtained by NUPGA is less than 0.1dB. In Fig.10 we notice that the gain is up  to 1.2dB, which indicates that the NUPGA is advantageous for low rates. In the comparison shown in Fig.11, we observe that the NUPGA extension algorithm achieves a performance improvement over that obtained by the NUPGA shortening algorithm, which is around 0.1 dB. We can see that according to the curves a little incremental extension has good BER performance, maintaining the same FER performance as the original code \cite{Vangala}. Future work will consider the application of NUPGA to detection problems  \cite{wence,mmimo,jidf,deLamare2003,itic,deLamare2008,cai2009,jiomimo,dfcc,deLamare2013,did,rrmser,1bitidd,aaidd,dynovs,memd,bfpeg,dopeg,vfap}.

\begin{figure}[htb]
\begin{center}
\includegraphics[scale=0.45]{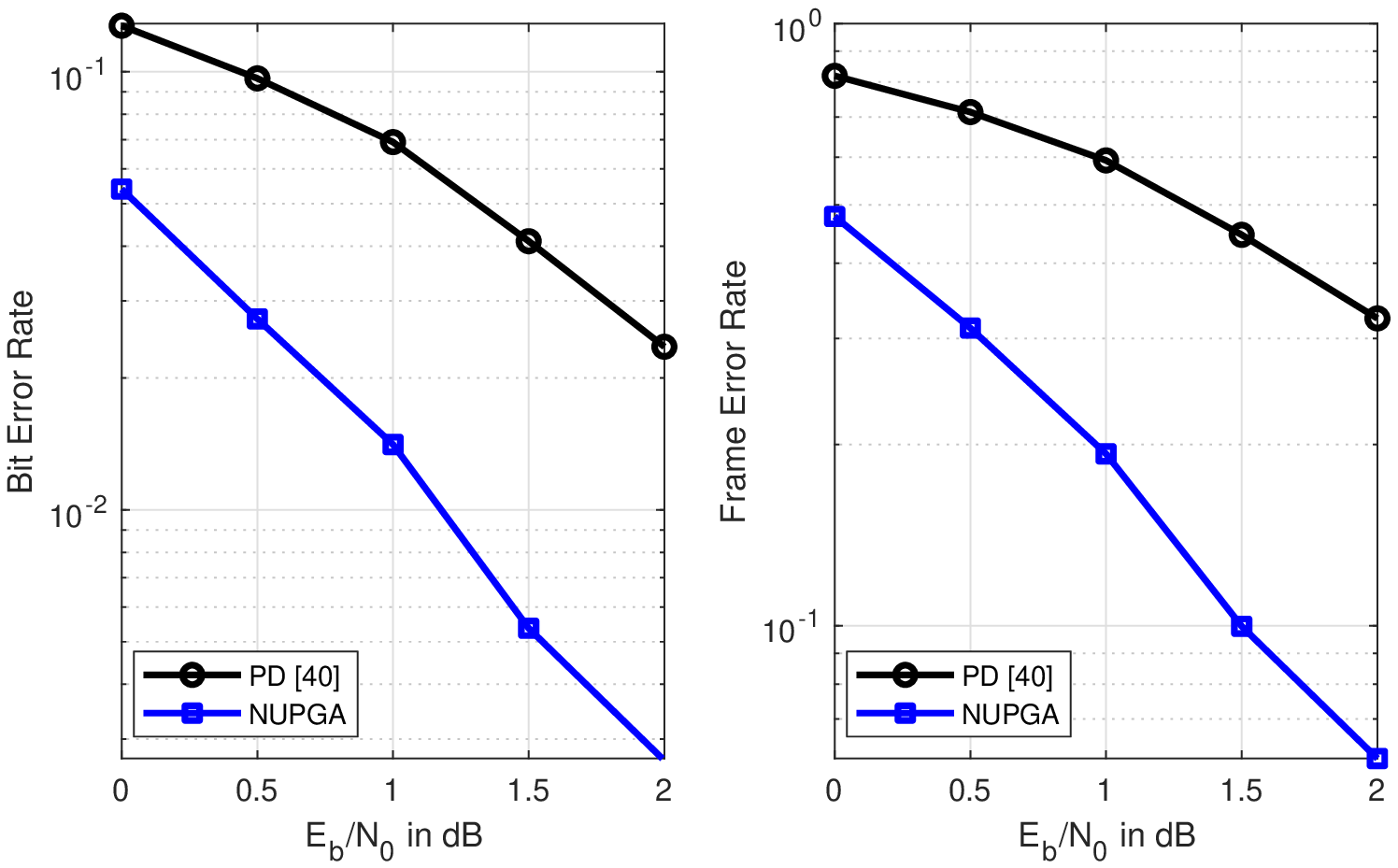}
\vspace{-0.5 em}
\caption{Performance of PD and NUPGA shortening algorithms both with $M=400$ and $K=50$ using CA-SCL with $L=16$ and $CRC=24$.}
\end{center}
\label{figura:fig_400_50_L16_C24}
\vspace{-0.5 em}
\end{figure}

\begin{figure}[htb]
\vspace{-1 em}
\begin{center}
\includegraphics[scale=0.45]{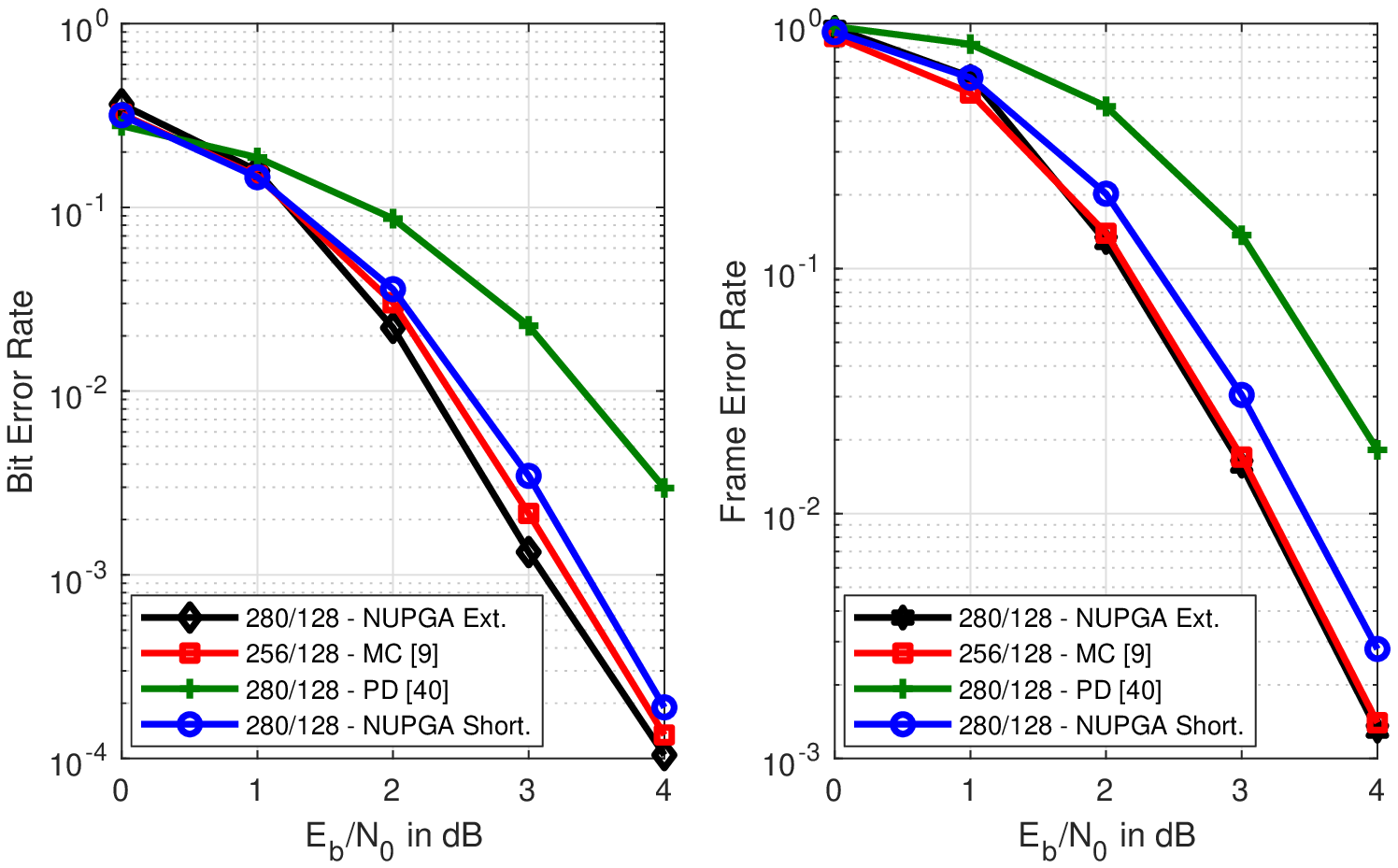}
\vspace{-0.5 em}
\caption{Performance of NUPGA extension algorithm, PD and NUPGA shortening algorithms using CA-SCL with $L = 16$ and CRC with length $24$.}
\end{center}
\label{figura:fig_256_128_S_E_4}
\vspace{-1 em}
\end{figure}

\section{Conclusions}

In this work, we have proposed a rate-compatible scheme for
constructing arbitrary-length polar codes using a non-uniform
channel polarization technique. We have then developed NUPGA-based
code design algorithms that exploit the non-uniform channel
polarization and the GA technique along with a proof that it
achieves capacity. With the proposed NUPGA-based shortening and
extension algorithms we can design polar codes for any code length
by re-polarizing the length of a conventional polar code. This is in
contrast to common schemes like punctured or shortened polar codes.
Since the performance of these length-compatible polar codes
(including extended, punctured or shortened) deteriorates with the
distance to code length of the mother code, our scheme considerably
outperforms existing schemes. Simulations illustrate the excellent
performance of NUPGA-based designs for short codes.









\begin{thebibliography}{99}

\bibitem{Arikan}
E. Arikan, ``Channel Polarization: A Method for Constructing
Capacity-Achieving Codes for Symmetric Binary-Input Memoryless
Channels", \textit{IEEE Transactions on Information Theory}, vol. 55
no. 7, pp. 3051-3073, July 2009.

\bibitem{5G}
(2018). ``Technical Specification Group Radio Access Network" [Online]. Available: \textit{http://www.3gpp.org/ftp/Specs/archive/38series/38.212/}

\bibitem{Mori1}
R. Mori and T. Tanaka, ``Performance of polar codes with the construction using density evolution", \textit{IEEE Communications Letters}, vol. 13, no. 7, pp. 519–521, July 2009.

\bibitem{Mori2}
R. Mori and T. Tanaka, ``Performance and construction of polar codes on symmetric binary-input memoryless channels", \textit{International Symposium on Information Theory (ISIT)}, 2009, pp. 1496–1500.

\bibitem{Tal1}
I. Tal and A. Vardy, ``How to Construct polar codes", \textit{IEEE Transactions on Information Theory}, vol. 59, no. 10, pp. 6562-6582, 2013.

\bibitem{Trifonov}
P. Trifonov, ``Efficient design and decoding of polar codes,” \textit{IEEE Transactions on Communications}, vol. 60, no. 11, pp. 1–7, 2012.

\bibitem{RZhang}
Y. Ge, R. Zhang, W. Tong, ``B-expansion: A Theoretical Framework for Fast and Recursive Construction of Polar Codes", \textit{IEEE Global Communication Conference}, 2017.

\bibitem{Schurch}
C. Schurch, ``A partial Order for the Synthesized Channels of a Polar Code", \textit{IEEE International Symposium on Information Theory (ISIT)}, pp. 220–224, 2016.

\bibitem{Vangala}
H. Vangala, E. Viterbo and Y. Hong, ``A Comparative Study of Polar Code Constructions for the AWGN Channel", \textit{https://arxiv.org/pdf/1501.02473.pdf}, Jan 2015.

\bibitem{Cheng}
J. Li, M. Hu and Z. Cheng, ``Research on Polar Code Construction Algorithms under Gaussian Channel", \textit{2018 Tenth International Conference on Ubiquitous and Future Networks (ICUFN)}, July 2018.

\bibitem{Wu}
D. Wu, Y. Li and Y. Sun, ``Construction and Block Error Rate Analysis of Polar Codes Over AWGN Channel Based on Gaussian Approximation", \textit{IEEE Communication Letters}, vol. 18, no. 7, Jul 2014.

\bibitem{Yuan}
H. Li and J. Yuan, ``A pratical construction method for Polar Codes in AWGN channels", \textit{IEEE 2013 Tencon - Spring}, April 2013.

\bibitem{P. Yuan}
P. Yuan, T. Prinz, and G. Bocherer, ``Polar Code Construction for List Decoding", \textit{ArXiv e-prints}, July 2017.

\bibitem{M. Qin}
M. Qin, J. Guo, A. Bhatia, A. G. i Fabregas, and P. Siegel, ``Polar Code Constructions Based on LLR Evolution", \textit{IEEE Commun. Lett.}, vol. 21, no. 6, pp. 1221–1224, June 2017.

\bibitem{S. Sun}
S. Sun and Z. Zhang, ``Designing Practical Polar Codes Using Simulation-Based Bit Selection", \textit{IEEE J. Emerging and Sel. Topics Circuits Syst.}, vol. 7, no. 4, pp. 594–603, Dec. 2017.

\bibitem{A. Elkelesh}
A. Elkelesh, M. Ebada, S. Cammerer and S. ten Brink, ``Genetic Algorithm-based Polar Code Construction for the AWGN Channel", \textit{SCC 2019; 12th International ITG Conference on Systems, Communications and Coding}, March 2019.

\bibitem{S.B. Korada}
S.B. Korada, E. Sasoglu, and R. Urbanke, ``Polar codes: Characterization of exponent, bounds, and constructions", \textit{IEEE Trans. Inf. Theory}, vol.56, no.12, pp.6253–6264, 2010.

\bibitem{N. Presman}
N. Presman, O. Shapira, and S. Litsyn, ``Binary polar code kernels
from code decompositions", \textit{Proc. IEEE Int. Symp. Inform. Theory (ISIT)}, pp.179–183, July 2011.

\bibitem{LZhang}
L. Zhang, Z. Zhang, and X. Wang, ``Polar code with block-length $N=3^n$", \textit{Wireless Communications and Signal Processing (WCSP), International Conference}, pp. 1–6, October 2012.

\bibitem{Serbetc}
B. Serbetci and A. E. Pusane, ``Practical polar code construction using generalised generator matrices", \textit{IET Communications}, vol. 8, no. 4, pp.419–426, March 2014.

\bibitem{Zhiliang}
Zhiliang Huang at all, ``On the Successive Cancellation Decoding of Polar Codes with Arbitrary Linear Binary Kernel", \textit{https://arxiv.org/pdf/1701.03264v2.pdf}, Jan 2017.

\bibitem{Gabry}
F. Gabry, V. Bioglio, I. Land, and J.-C. Belfiore, ``Multi-Kernel Construction of Polar Codes", \textit{IEEE Int. Conf. Commun. (ICC)}, 2017.

\bibitem{Benammar}
M. Benammar, V. Bioglio, F. Gabry, and I. Land, ``Multi-Kernel Polar Codes: Proof of Polarization and Error Exponents", \textit{IEEE Info. Theory Workshop}, 2017.

\bibitem{Hussami}
N. Hussami, S. B. Korada, and R. Urbanke, ``Performance of Polar Codes for Channel and Source Coding", \textit{IEEE Inter. Symp. Inf. Theory (ISIT)}, June 2009, pp. 1488–1492.

\bibitem{B. Li}
B. Li, H. Shen, and D. Tse, ``A RM-Polar Codes", \textit{ArXiv e-prints}, July 2014.

\bibitem{M. Mondelli}
M. Mondelli, S. H. Hassani, and R. L. Urbanke, ``From Polar to ReedMuller Codes: A Technique to Improve the Finite-Length Performance", \textit{IEEE Trans. Commun.}, vol. 62, no. 9, pp. 3084–3091, Sep. 2014.

\bibitem{Bioglio}
V. Bioglio, F. Gabry and I. Land, ``Low-Complexity Puncturing and Shortening of Polar Codes", \textit{2017 IEEE Wireless Communications and Networking Conference Workshops (WCNCW)}, San Francisco, CA, pp. 1-6, 2017.

\bibitem{Hong}
S. N. Hong et al., ``Capacity-achieving rate-compatible polar
codes", \textit{arXiv preprint arXiv:1510.01776}, 2015.

\bibitem{Saber}
H. Saber and I. Marsland, ``An incremental redundancy Hybrid ARQ
scheme via puncturing and extending of polar codes", \textit{IEEE
Transactions on Communications}, vol. 63, no. 11, pp. 3964$-$3973,
2015.

\bibitem{Chen}
K. Chen, K. Niu and J. Lin, ``A hybrid ARQ scheme based on polar
codes", \textit{IEEE Communications Letters}, vol. 17, no. 10, pp.
1996$-$1999, 2013.

\bibitem{NiuChen}
K. Niu, K. Chen and J.-R. Lin, ``Beyond turbo codes: rate-compatible
punctured polar codes", in \textit{IEEE International Conference on
Communications (ICC)}, 2013, pp. 3423-3427.

\bibitem{Shin}
D. Shin et al., ``Design of Length-Compatible Polar Codes Based on
the Reduction of Polarizing Matrices", \textit{IEEE Transactions on
Communications}, vol. 61, no. 7, July 2013.

\bibitem{Eslami}
A. Eslami and H. Pishro-Nik, ``A practical approach to polar codes",
in \textit{IEEE International Symposium on Information Theory
Proceedings(ISIT)}, 2011, pp. 16-20.

\bibitem{EslamiPishro}
A. Eslami and H. Pishro-Nik, ``On finite-length performance of polar
codes: stopping sets, error floor, and concatenated design",
\textit{IEEE Transactions on Communications}, vol. 61, no. 3, March
2013.

\bibitem{Kim}
J. Kim et al., ``An Efficient Search on Puncturing Patterns for
Short Polar Codes", \textit{International Conference on Information
and Communication Technology Convergence (ICTC)}, 2015.

\bibitem{Zhang}
L. Zhang et al., ``On the Puncturing Patterns for Punctured Polar
Codes", \textit{IEEE International Symposium on Information Theory},
2014.

\bibitem{Miloslavskaya}
V. Miloslavskaya, ``Shortened polar codes", \textit{IEEE
Transactions on Information Theory}, vol. 61, no. 9, pp. 4852-4865,
2015.

\bibitem{Wang}
R. Wang and R. Liu, ``A novel puncturing scheme for polar codes",
\textit{IEEE Communications Letters}, vol. 18, no. 12, pp.
2081-2084, 2014.

\bibitem{Niu}
K. Niu et al., ``Rate-Compatible Punctured Polar Codes: Optimal
Construction Based on Polar Spectra",
\textit{https://arxiv.org/pdf/1612.01352}, 2016.

\bibitem{Oliveira}
R. M. Oliveira and R. C. de Lamare, ``Rate-Compatible Polar Codes Based on Polarization-Driven Shortening", \textit{IEEE Communications Letters}, vol. 22, no. 10, pp. 1984-1987, 2018.

\bibitem{MJang}
M. Jang et al., ``Rate matching for polar codes based on binary domination", \textit{IEEE Transactions on Communications}, vol. 67, no. 10, pp. 6668-6681, 2019.

\bibitem{Saber}
H. Saber and I. Marsland, ``An Incremental Redundancy Hybrid ARQ Scheme via Puncturing and Extending of Polar Codes", \textit{IEEE Transactions on Communications}, vol. 63, no. 11, pp. 3964-3973, 2015.

\bibitem{MZhao}
M. Zhao at all, ``An Adaptive IR-HARQ Scheme for Polar Codes by Polarizing Matrix Extension", \textit{IEEE Communications Letters}, vol. 22, no. 7, pp. 1306-1309, 2018.

\bibitem{Huang}
Yu-Ming Huang at all, ``Re-Polarization Processing in Extended Polar Codes", \textit{IEICE Transactions on Communications }, vol. E100-B, no. 10, pp. 1765-1777, 2017.

\bibitem{PTrifonov}
P. Trifonov and P. Semenov, ``Generalized concatenated codes based on polar codes", \textit{8th International Symposium on Wireless Communication Systems}, 2011.

\bibitem{Mahdavifar}
H. Mahdavifar at all, ``On the construction and decoding of concatenated polar codes", \textit{International Symposium on Information Theory (ISIT)}, pp. 952-956, July 2013.

\bibitem{Cavatassi}
A. Cavatassi at all, ``Asymmetric Construction of Low-Latency and Length-Flexible Polar Codes", \textit{IEEE International Conference on Communications (ICC)}, July 2019.

\bibitem{Trifonov2}
P. Trifonov, ``Chained Polar Subcodes", \textit{11th International ITG Conference on Systems, Communications and Coding (SCC)}, June 2017.

\bibitem{Mahdavifar}
H. Mahdavifar, M. El-Khamy, J. Lee and I. Kang, ``Compound polar codes", \textit{2013 Information Theory and Applications Workshop (ITA)}, pp. 1-6, 2013.

\bibitem{Kim2}
J. Kim, J. Lee, ``Polar codes for non-identically distributed channels", \textit{EURASIP Journal on Wireless Communications and
Networking}, 2016:287, 2016.

\bibitem{Leroux}
C.Leroux, I.Tal, A.Vardy,and W.J. Gross, ``Hardware architectures
for successive cancellation decoding of polar codes", \textit{IEEE
ICASSP}, pp. 1665-1668, 2011.

\bibitem{Tal2}
I. Tal and A. Vardy, ``List Decoding of Polar Codes", \textit{IEEE Transactions on Information Theory}, vol. 61, no. 5, pp. 2213-2226, May 2015.

\bibitem{Oliveira2}
R. M. Oliveira and R. C. de Lamare, ``Non-Uniform Channel Polarization and Design of Rate-Compatible Polar Codes", \textit{16th International Symposium on Wireless Communication Systems (ISWCS)}, Oulu, Finland, pp. 537-541, 2019.

\bibitem{nupga_access}
R. M. Oliveira and R. C. de Lamare, ``Design of Rate-Compatible Polar Codes Based on Non-Uniform Channel Polarization", IEEE Access, 2021.

\bibitem{mmimo}
R. C. de Lamare, "Massive MIMO systems: Signal processing challenges
and future trends," URSI Radio Science Bulletin, vol. 2013, no. 347,
pp. 8-20, Dec. 2013.

\bibitem{wence}
W. Zhang et al., "Large-Scale Antenna Systems With UL/DL Hardware
Mismatch: Achievable Rates Analysis and Calibration," IEEE
Transactions on Communications, vol. 63, no. 4, pp. 1216-1229, April
2015.

\bibitem{Li2011}
P. Li, R. C. de Lamare and R. Fa, ``Multiple Feedback Successive
Interference Cancellation Detection for Multiuser MIMO Systems",
IEEE Trans. on Wireless Comm., vol. 10, no. 8, pp. 2434-2439, Aug.
2011.

%

\bibitem{deLamare2003}
R. C. de Lamare and R. Sampaio-Neto, "Adaptive MBER decision
feedback multiuser receivers in frequency selective fading
channels," IEEE Communications Letters, vol. 7, no. 2, pp. 73-75,
Feb. 2003.

\bibitem{itic}
R. C. De Lamare, R. Sampaio-Neto and A. Hjorungnes, "Joint iterative
interference cancellation and parameter estimation for cdma
systems," IEEE Communications Letters, vol. 11, no. 12, pp. 916-918,
December 2007.

\bibitem{deLamare2008}
R. C. De Lamare and R. Sampaio-Neto, ``Minimum Mean-Squared Error
Iterative Successive Parallel Arbitrated Decision Feedback Detectors
for DS-CDMA Systems,'' IEEE Transactions on Communications, vol. 56,
no. 5, pp. 778-789, May 2008.

%
\bibitem{cai2009}
Y. Cai and R. C. de Lamare, "Space-Time Adaptive MMSE Multiuser
Decision Feedback Detectors With Multiple-Feedback Interference
Cancellation for CDMA Systems," IEEE Transactions on Vehicular
Technology, vol. 58, no. 8, pp. 4129-4140, Oct. 2009.

\bibitem{jiomimo}
R. C. de Lamare and R. Sampaio-Neto, "Adaptive Reduced-Rank
Equalization Algorithms Based on Alternating Optimization Design
Techniques for MIMO Systems," IEEE Transactions on Vehicular
Technology, vol. 60, no. 6, pp. 2482-2494, July 2011.

\bibitem{dfcc}
P. Li and R. C. De Lamare, "Adaptive Decision-Feedback Detection
With Constellation Constraints for MIMO Systems," IEEE Transactions
on Vehicular Technology, vol. 61, no. 2, pp. 853-859, Feb. 2012.

\bibitem{deLamare2013}
R. C. de Lamare, ``Adaptive and Iterative Multi-Branch MMSE Decision
Feedback Detection Algorithms for Multi-Antenna Systems,'' IEEE
Transactions on Wireless Communications, vol. 12, no. 10, pp.
5294-5308, October 2013.

\bibitem{did}
P. Li and R. C. de Lamare, "Distributed Iterative Detection With
Reduced Message Passing for Networked MIMO Cellular Systems,"  IEEE
Transactions on Vehicular Technology, vol. 63, no. 6, pp. 2947-2954,
July 2014.

\bibitem{rrmser}
Y. Cai, R. C. de Lamare, B. Champagne, B. Qin and M. Zhao, "Adaptive
Reduced-Rank Receive Processing Based on Minimum Symbol-Error-Rate
Criterion for Large-Scale Multiple-Antenna Systems," IEEE
Transactions on Communications, vol. 63, no. 11, pp. 4185-4201, Nov.
2015.

\bibitem{jidf}
R. C. de Lamare and R. Sampaio-Neto, "Adaptive Reduced-Rank
Processing Based on Joint and Iterative Interpolation, Decimation,
and Filtering," in IEEE Transactions on Signal Processing, vol. 57,
no. 7, pp. 2503-2514, July 2009.

\bibitem{Angelosante2010}
D. Angelosante, J. A. Bazerque and G. B. Giannakis, ``Online
adaptive estimation of sparse signals: where RLS meets the
$l_1$-norm,'' IEEE Trans. Sig. Proc., vol. 58, no. 7, pp. 3436-3446,
2010.

\bibitem{l1stap} Z. Yang, R. C. de Lamare and X. Li, "$L_1$
-Regularized STAP Algorithms With a Generalized Sidelobe Canceler
Architecture for Airborne Radar," in IEEE Transactions on Signal
Processing, vol. 60, no. 2, pp. 674-686, Feb. 2012.

\bibitem{saalt}
R. C. de Lamare and R. Sampaio-Neto, "Sparsity-Aware Adaptive
Algorithms Based on Alternating Optimization and Shrinkage," in IEEE
Signal Processing Letters, vol. 21, no. 2, pp. 225-229, Feb. 2014.


%
\bibitem{JChoi2005}
Jihoon Choi et. al, ``Adaptive MIMO decision feedback equalization for receivers with time-varying channels,'' in \textit{IEEE Trans. on Sig. Proc.}, vol. 53, no. 11, pp. 4295-4303, Nov. 2005.

%
\bibitem{Bradley98}
P. S. Bradley et. al, ``Feature selection via concave minimization and support vector machines,'' in \textit{Proc. 13th ICML}, 1998, pp. 82â€“90.

%
\bibitem{XWang1999}
Xiaodong Wang and H. V. Poor, ``Iterative (turbo) soft interference cancellation and decoding for coded CDMA,'' in \textit{IEEE Trans. on Comm.}, vol. 47, no. 7, pp. 1046-1061, July 1999.

\bibitem{bfidd}
A. G. D. Uchoa, C. T. Healy and R. C. de Lamare, "Iterative
Detection and Decoding Algorithms for MIMO Systems in Block-Fading
Channels Using LDPC Codes," IEEE Transactions on Vehicular
Technology, vol. 65, no. 4, pp. 2735-2741, April 2016.

\bibitem{1bitidd}
Z. Shao, R. C. de Lamare and L. T. N. Landau, "Iterative Detection
and Decoding for Large-Scale Multiple-Antenna Systems With 1-Bit
ADCs," IEEE Wireless Communications Letters, vol. 7, no. 3, pp.
476-479, June 2018.

\bibitem{aaidd}
R. B. di Renna and R. C. de Lamare, ``Adaptive Activity-Aware
Iterative Detection for Massive Machine-Type Communications", IEEE
Wireless Communications Letters, 2019.

\bibitem{dynovs}
Z. Shao, L. T. N. Landau and R. C. de Lamare, ``Dynamic Oversampling for 1-Bit ADCs in Large-Scale Multiple-Antenna Systems," in IEEE Transactions on Communications

\bibitem{bfpeg}
A. G. D. Uchoa, C. Healy, R. C. de Lamare and R. D. Souza, ``Design of LDPC Codes Based on Progressive Edge Growth Techniques for Block Fading Channels," in IEEE Communications Letters, vol. 15, no. 11, pp. 1221-1223, November 2011

\bibitem{dopeg} C. T. Healy and R. C. de Lamare,
"Decoder-Optimised Progressive Edge Growth Algorithms for the Design
of LDPC Codes with Low Error Floors," in IEEE Communications
Letters, vol. 16, no. 6, pp. 889-892, June 2012.

\bibitem{memd}
C. T. Healy and R. C. de Lamare, "Design of LDPC Codes Based on
Multipath EMD Strategies for Progressive Edge Growth," in IEEE
Transactions on Communications, vol. 64, no. 8, pp. 3208-3219, Aug.
2016.

\bibitem{vfap}
J. Liu and R. C. de Lamare, "Low-Latency Reweighted Belief
Propagation Decoding for LDPC Codes," IEEE Communications Letters,
vol. 16, no. 10, pp. 1660-1663, October 2012.

\bibitem{LLiu2018}
L. Liu et. al, ``Sparse Signal Processing for Grant-Free Massive Connectivity: A Future Paradigm for Random Access Protocols in the Internet of Things,'' in \textit{IEEE Sig. Proc. Mag.}, vol. 35, no. 5, pp. 88-99, Sept. 2018.

\end{thebibliography}
\end{document}